OPEN

# Simulating high-speed solar wind streams from coronal holes using an L5-L1 configuration of Lagrangian points

Tatiana Podladchikova[1,2]✉, Astrid M. Veronig[2,3], Manuela Temmer[2] & Stefan J. Hofmeister[4]

**Coronal holes (CHs) are known to be sources of high-speed solar wind streams (HSSs), yet the physical mechanisms linking CH position and characteristics to solar wind (SW) behaviour remain unclear. Our results reveal that the latitude of CHs, especially smaller ones, combined with the heliographic latitude of the solar disk's central point (B0 angle), plays a critical role in driving discrepancies in SW velocity across the heliosphere. To investigate this, we use archival data from STEREO-B, STEREO-A, and Earth to simulate an L5-L1 configuration, where L5 is a vantage point approximately 60° behind Earth in its orbit (as proposed for the Vigil mission), and L1 is between Earth and the Sun where SW measurements are typically taken. We use these insights to develop a predictive algorithm for HSSs, beginning with an analysis of the separation angle and distances between L5 and L1. We then introduce a predictive indicator and empirical criteria based on CH properties and the B0 angle to adjust for changes in SW velocity at L1. Our results show that the L5 viewpoint demonstrates the capability to significantly improve the accuracy and lead times of HSS predictions, enhancing our understanding of the CH-HSS relationship and potentially improving space weather forecasting.**

The heliosphere, the vast bubble surrounding the solar system, is shaped by the solar wind (SW) - a continuous stream of charged particles emitted by the Sun at velocities ranging from 250 to 800 km/s. This SW not only transports the Sun's magnetic field but also shapes it into a spiral configuration as the Sun rotates, forming the interplanetary magnetic field (IMF)[1]. High-speed solar wind streams (HSSs) originate from coronal holes (CHs) in the Sun's atmosphere, where the Sun's magnetic field opens up, allowing ionised particles to escape. Visible as dark regions in extreme ultraviolet (EUV) and X-ray images, CHs are areas of reduced density and temperature, and are crucial for understanding how SW streams accelerate and propagate through the solar system. While HSSs do not cause the strongest geomagnetic storms -those are triggered by Earth-directed coronal mass ejections (CMEs) - they do generate more frequent medium-sized geomagnetic storms, potentially contributing more overall energy to the Earth system than the rarer but more intense CMEs[2–4].

Understanding the physical mechanisms that link CHs to the generation of HSSs remains a challenge, complicating the accuracy of HSS predictions. The flux tube expansion factor, introduced in the Wang-Sheeley (WS) model[5], connects the expansion of magnetic fields from CHs to SW speeds, with larger expansions resulting in slower winds. This model was later extended into the Wang-Sheeley-Arge (WSA) model, which incorporates magnetic field and solar surface data for more accurate forecasts[6]. Coronal and heliospheric MHD simulations[7–9], along with the ENLIL[10], the European heliospheric forecasting information asset (EUHFORIA)[11], and others[12–16], simulate 3D SW propagation in the heliosphere, have further refined these forecasts. Additionally, current forecasting methods largely rely on correlating CH area with HSS velocity to predict peak speeds up to five days in advance and anticipate associated geomagnetic storms[17–25]. Follow-up studies further explored how the velocity of HSSs measured at Earth varies with the location of the CH on the Sun, introducing a CH area-velocity relationship that is dependent on CH latitude[26–28]. However, forecasts typically rely on data from the L1 Lagrange point, positioned between the Earth and the Sun, which provides limited lead time and only a partial view of solar phenomena as they approach Earth.

[1]Skolkovo Institute of Science and Technology, 121205 Moscow, Russia. [2]Institute of Physics, University of Graz, 8010 Graz, Austria. [3]Kanzelhöhe Observatory for Solar and Environmental Research, University of Graz, 9521 Treffen, Austria. [4]Columbia University, 10027 New York, NY, USA. ✉email: t.podladchikova@skol.tech; tatiana.podladchikova@uni-graz.at

Several studies[23,29–32] have investigated the potential of using observations from the L5 Lagrange point-located about 60° behind Earth in its orbit-to improve SW forecasting at L1. This vantage point provides an early view of CHs before they rotate into Earth's line of sight, offering a valuable opportunity to extend lead times for predictions and to observe the evolution of CHs over time. However, these attempts assume a steady SW during the time it takes for structures to rotate between the L5 and L1 positions. A number of studies highlighted the need for caution when combining in-situ spacecraft data from different locations due to positional differences[33–36]. Given that HSSs from CHs are extended streams, the latitude at which a spacecraft samples the stream significantly affects the persistence-based forecasting of stream velocity[26,37].

In this study, we focus on enhancing our understanding of the mechanisms that drive HSSs by incorporating the L5–L1 observational configuration and CH data, as envisioned in the upcoming Vigil mission. By analysing data from Solar Terrestrial Relations Observatory (STEREO-B, STEREO-A), and Earth to simulate this configuration, we identify key factors-such as the latitude of CHs and the B0 angle, which represents the heliographic latitude of the solar disk centre (the heliographic latitude of a spacecraft) -that contribute to discrepancies in SW velocity measurements between L5 and L1. Leveraging these insights, we develop a novel predictive algorithm that refines initial forecasts by integrating these factors, thereby improving the accuracy and extending the lead times of SW predictions.

## Data to simulate an L5–L1 configuration

To simulate an L5–L1 configuration, we use 1-hour averaged data from STEREO-B, STEREO-A and OMNI-2[38] database for the period 2008–2010, which corresponds to separation angles of $22 - 90°$ between Earth and each STEREO spacecraft. We consider three combinations of spacecraft to model L5–L1 conditions: we predict from STEREO-B to STEREO-A, from STEREO-B to Earth, and from Earth to STEREO-A. We employ an operational tool at the University of Graz to detect CHs by analysing EUV 195 and 193 Å images from the STEREO-B/STEREO-A and Solar Dynamics Observatory (SDO) spacecraft, respectively, using a segmentation method based on intensity distribution[21]. We define the CH area as the fraction of the solar surface occupied by a CH within a $15°$-wide meridional slice centred on the solar central meridian, covering approximately one day of solar synodic rotation, and spanning latitudes from $-60$ to $60°$. The latitude of CH location is estimated in its "centre of masses"[39]. To cut out periods of coronal mass ejections (CME) occurrence from the SW profile, we use the information on interplanetary coronal mass ejections (ICME) observed with STEREO-A and STEREO-B as well as the Richardson & Cane ICME list for Earth.

## Methods and results

Our approach consists of three main steps. First, we perform initial predictions of velocities of HSSs using a longitudinal separation angle between each pair of probes and accounting for the distance from simulated "L5" and "L1" to the Sun. Second, we analyse the causes of the observed discrepancies in HSS velocity at "L5" and "L1", highlighting a new relationship and demonstrating that CH latitude, combined with the B0 angle, plays a crucial role in predicting whether an increase or decrease in SW velocity should be anticipated. Moreover, the effect on SW velocity - whether it increases or decreases - is more pronounced for smaller CH areas, higher CH latitudes, and higher B0 angles. Finally, we develop criteria for predicting whether HSSs will result in an increase or decrease in SW velocity and adjust initial predictions based on these estimates. To present our approach, we use data from 2008 with STEREO-A (simulating "L1") and STEREO-B (modelling "L5"), with separation angles ranging from 44 to $88°$. Results for other years and spacecraft combinations are provided in Supplementary Information.

### Initial predictions of HSS velocities based on the L5–L1 configuration

To begin, we establish event associations between CHs and HSSs by identifying successive individual peaks in the CH areas and SW velocity time series, selecting the closest CH area peak within a [-6, -2] day window relative to the observed SW velocity peak. Figs. 1 and 2 provide an overview of CH fractional areas (blue, left Y-axis) and associated HSS velocity (red, right Y-axis) for 2008, shown for STEREO-B (Fig. 1) and STEREO-A (Fig. 2). CH data were processed to remove outliers and linearly interpolated to fill data gaps (up to 2%). To reduce noise, both CH and SW data were smoothed using the optimised running mean. The algorithm optimizes between two inherently conflicting criteria: data fidelity, which reflects the closeness of the approximating curve to the data, and the smoothness of the approximating curve.[40]. Each CH is labelled with a number on the plot, with the associated HSS marked by the same number. Red dashed vertical lines indicate the start and end of an ICME.

To derive the time shift $\Delta T$( 3.3–6.7 days, blue line in Fig. 3a, right Y-axis) for SW at "L5" (STEREO-B) to appear at "L1" (STEREO-A), we use the longitudinal separation angle $\alpha$ between the two spacecraft (44–88°, blue line in Fig. 3a, left Y-axis) to calculate

$$\Delta T = \frac{\alpha}{\Omega}, \tag{1}$$

with $\Omega$ denoting the synodic rotation rate, given as $\Omega = \frac{360}{27.27}$ $[°/\text{day}]$.

As illustrated in Fig. 3b, the distance from STEREO-B and STEREO-A to the Sun varies significantly throughout the year. Consequently, the arrival times of SW at STEREO-B and STEREO-A can differ depending on their respective distances from the Sun[33]. To better estimate when a given SW stream, originating from the same source region on the Sun, will be observed at the "L1" point (STEREO-A), we adjust the initial time shift based on when it was first observed at "L5" (STEREO-B) (see Fig. 3a, red line) using the following relation:

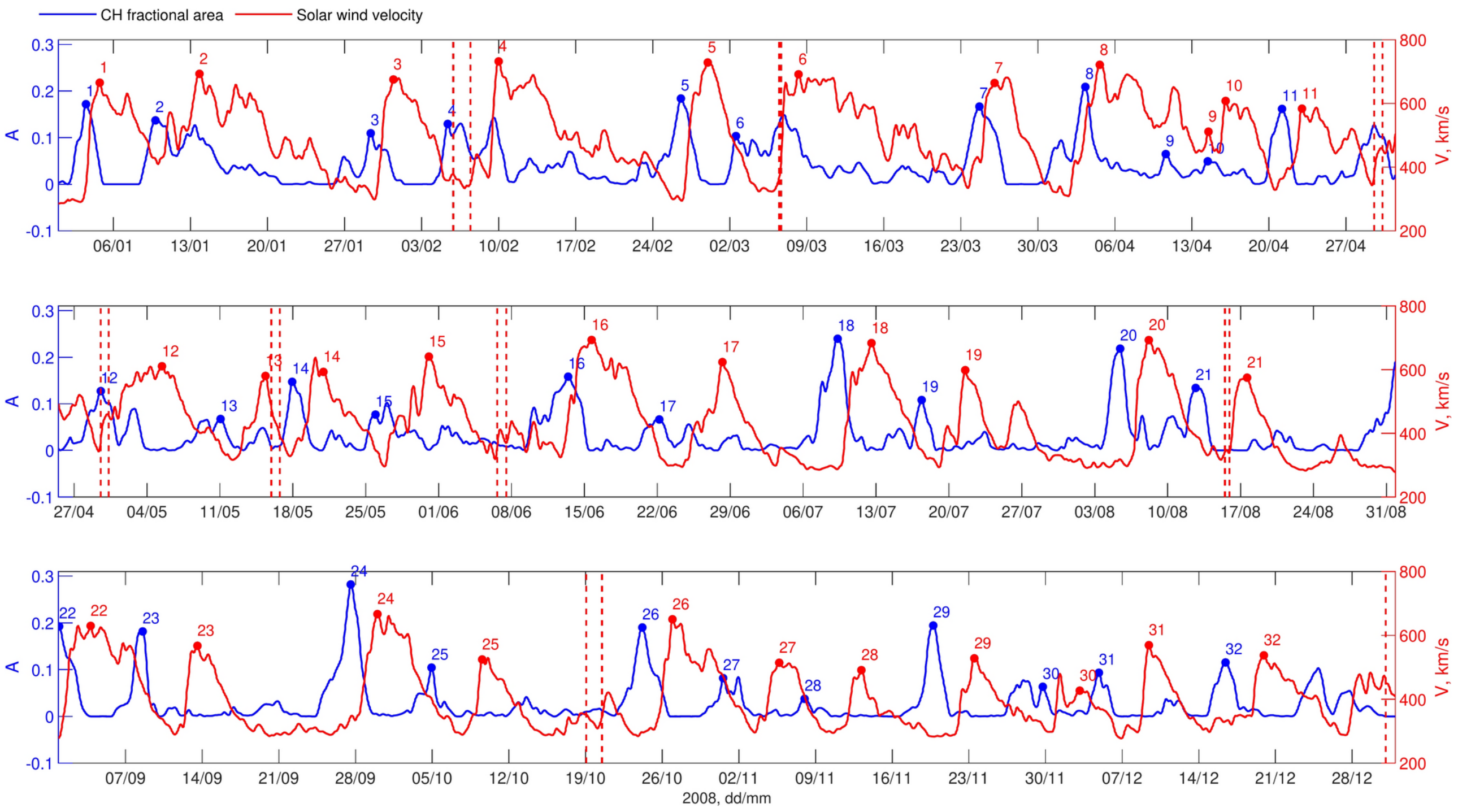

**Fig. 1**. CH fractional areas and associated HSS velocity for 2008 observed with STEREO-B("L5"). Numbers and markers shows associated CH-HSS pair. Red dashed vertical lines indicate start and end of an ICME.

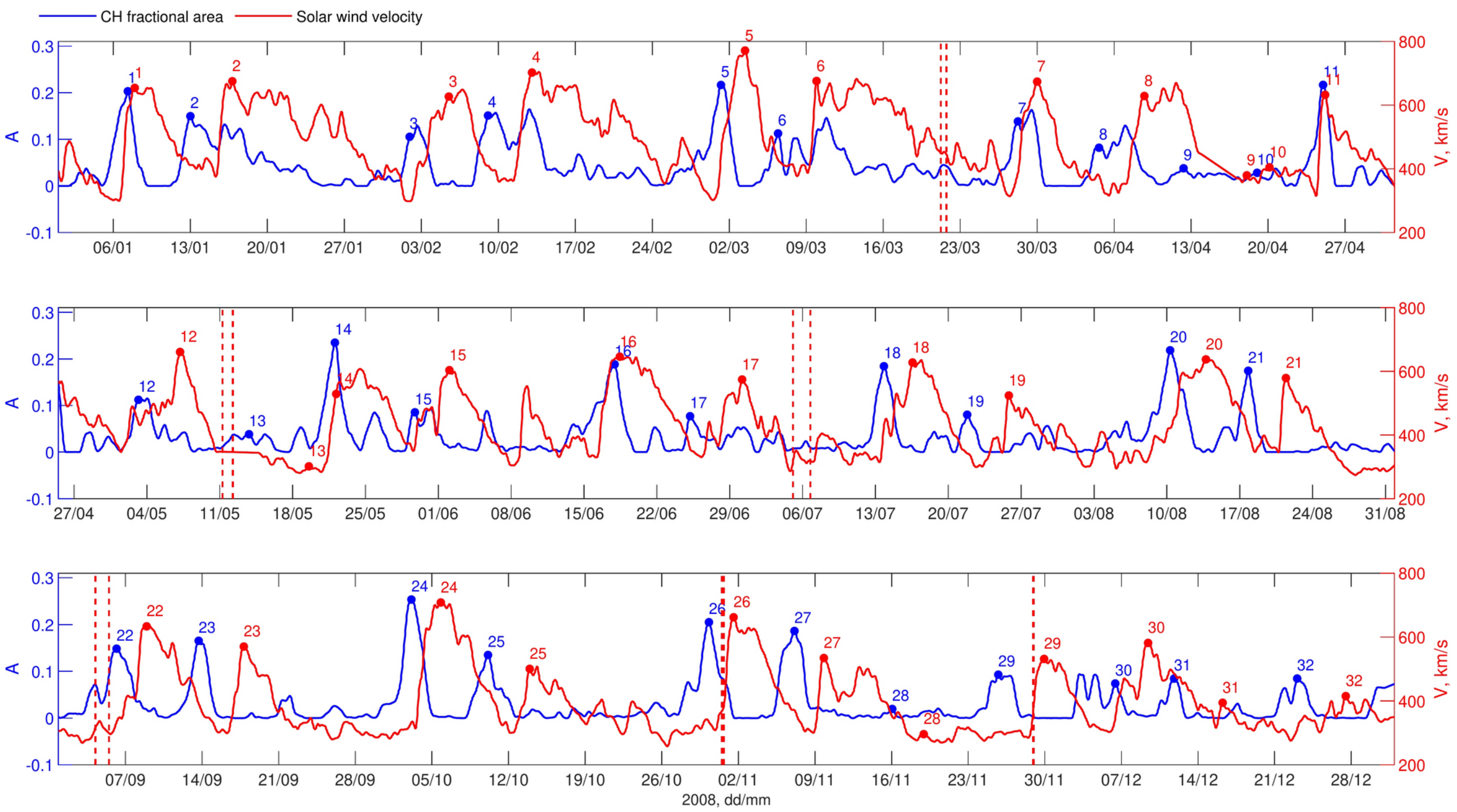

**Fig. 2**. The same, as Fig. 1, but for STEREO-A ("L1").

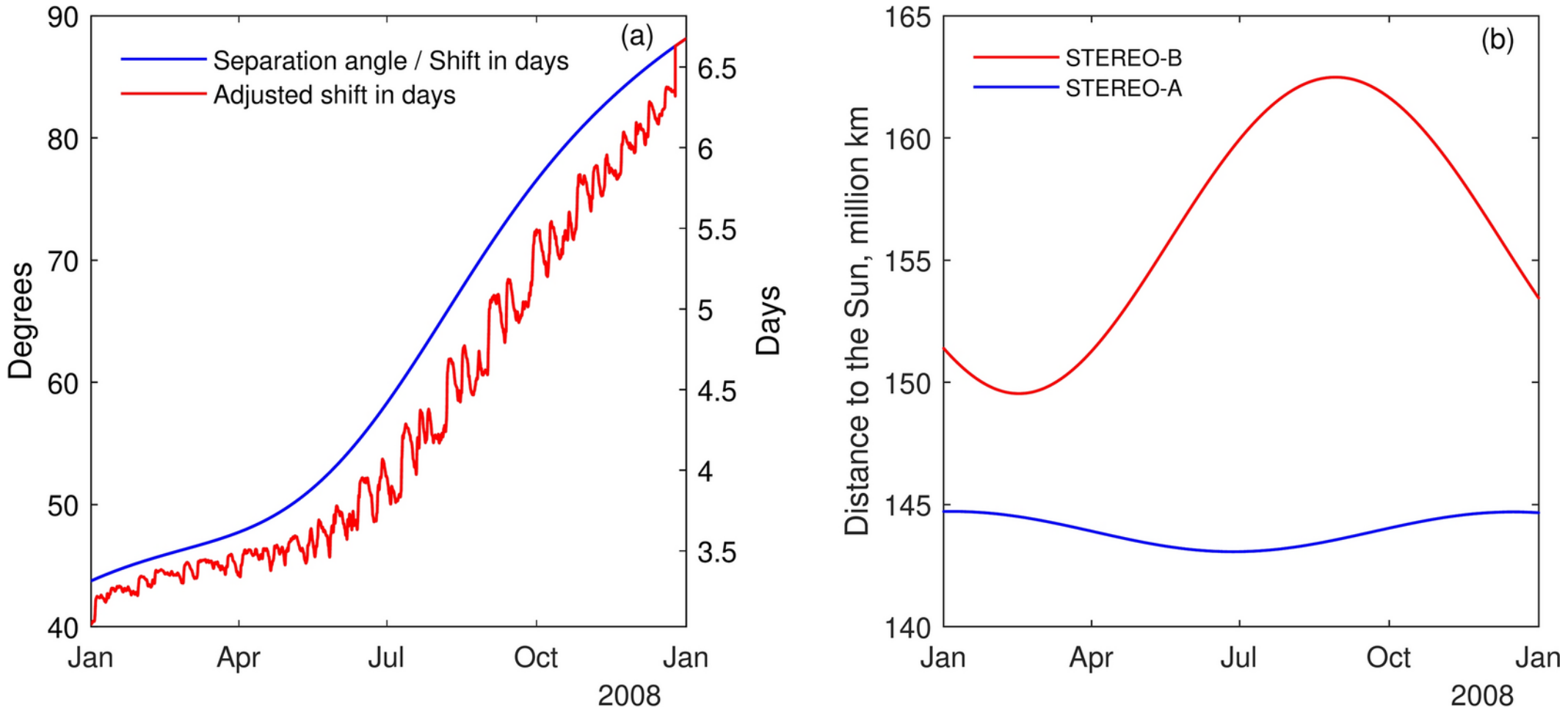

**Fig. 3**. The separation angle between STEREO-B and STEREO-A (panel a, blue line, left Y-axis) along with the associated time shift (in days, panel **a**, blue line, right Y-axis) for SW at "L5" (STEREO-B) to appear at "L1" (STEREO-A). Red line in panel **a** (right Y-axis) indicates the adjusted shift in days, accounting for the varying distance from STEREO-B and STEREO-A to the Sun in 2008 (panel **b**).

$$T_i = \frac{D_i^B - D_j^A}{V_i^B}. \tag{2}$$

Here $D_i^B$ is the distance from STEREO-B to the Sun at time *i*, $D_j^A$ is the corresponding distance from STEREO-A to the Sun at time *j*, defined from the the separation angle $\alpha$ between the two spacecraft, and $V_i^B$ is the SW velocity at time *i*. $T_i$ is subtracted from the originally derived time shift $\Delta T$( Fig. 3a, blue line, right Y-axis) and the results are shown by the red line in Fig. 3a (right Y-axis).

Figure 4 presents the prediction of SW velocity at "L1" (STEREO-A, blue, left Y-axis) using data from "L5" (STEREO-B, red, left Y-axis). The root mean square (RMS) of the prediction error is 75.4 km/s, with a correlation coefficient between the predicted and measured SW velocity of $r = 0.78$. Although the prediction accuracy is reasonably high, with prediction leads times ranging from 3.04 to 6.67 days for these specific conditions, Fig. 4 shows that for several HSSs the velocity peak predicted for STEREO-A is either overestimated or underestimated (e.g., No. 10, 13, 28, 30, 31, 32). In the next section, we analyse the causes of discrepancies in SW velocity at "L1" relative to "L5", incorporating CH information along with the B0 angle.

## Coronal hole and B0 angle effects on solar wind velocity discrepancies between "L1" and "L5"

As is known, the B0 angle can affect the apparent velocity of the SW and may contribute to differences in SW velocity measured at "L1" relative to "L5"[37]. The right Y-axis in Fig. 4 shows the B0 angle (magenta for STEREO-B, cyan for STEREO-A). Figure 4 reveals significant discrepancies between SW measurements at STEREO-B and STEREO-A during periods of large differences in the B0 angle between the two spacecraft. For instance, HSS No. 28 shows a strong overestimation (HSS missing at "L1", more than 50% relative difference between "L5" and "L1" SW velocity peaks), HSS No. 29 is relatively accurate (less than 10% difference), HSS No. 30 is an example of underestimation (HSS observed at both "L5" and "L1", less then 50% relative difference), and HSS No. 31 again shows overestimation (the same criteria). Similarly, HSS No. 13 represents a strong overestimation, while HSS No. 14 demonstrates a relatively accurate forecast. Thus, information about the B0 angle alone is insufficient to determine whether each individual case will result in overestimation or underestimation.

Figure 5 shows examples of CH-HSS pairs where the SW velocity forecast is either overestimated (top panels) or underestimated (bottom panels). Left panels show 195 Å STEREO-B images together with segmented CHs and $15^\circ$-wide meridional slice for 11 May 2008, 00:06 UT (top), and for 29 November 2008, 18:06 UT (bottom). In the right panels, the left Y-axis represents the CH area values for the"L5" (STEREO-B, dashed red line), and the right Y-axis indicates the corresponding SW velocities for "L5" (STEREO-B, solid red line) and "L1" (STEREO-A, solid blue line). Blue and red markers highlight the peaks of the associated CH-HSS pairs. Note that in this figure, the STEREO-A data are shifted back to the time of STEREO-B based on the estimated adjusted time shift (Fig. 3a, red line, right Y-axis).

For the overestimated prediction, the solar wind velocity at "L5" (STEREO-B) reaches a peak of 580 km/s (red). Meanwhile, the high-speed solar wind stream, shown in blue, mostly missed STEREO-A (blue). For this case, the CH latitude at "L5" (STEREO-B), $Lat^{L5} = -51^\circ$, is negative, indicating that the CH is located closer to high latitudes. The B0 angle at "L5" (STEREO-B), $B_0^{L5} = -5.7^\circ$, is also negative, though relatively large. In contrast, the B0 angle at "L1" (STEREO-A), $B_0^{L1} = 0.8^\circ$, is positive, creating a difference of more than $6^\circ$ between $B_0^{L5}$ and $B_0^{L1}$. This combination of parameters may indicate that the fast SW observed at "L5" (STEREO-B) is expected to be reduced by the time it reaches "L1" (STEREO-A).

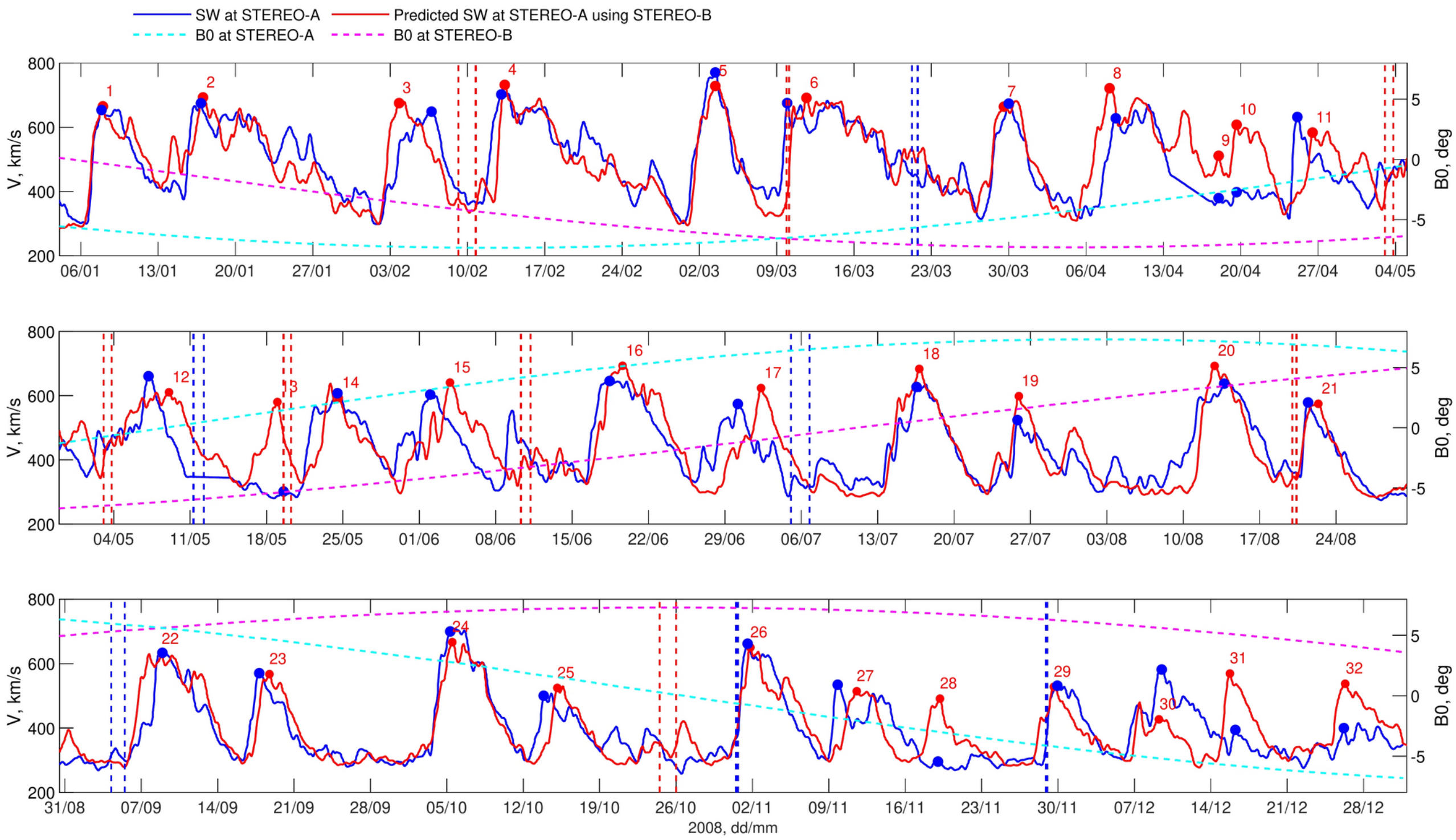

**Fig. 4**. Initial prediction of SW velocity (left Y-axis) at "L1" (STEREO-A, blue) in 2008 using data from "L5" (STEREO-B, red). The prediction, which takes into account the distance of both spacecraft from the Sun, is presented along with the B0 angle, indicated on the right Y-axis (magenta for STEREO-B, cyan for STEREO-A). Red (STEREO-B) and blue (STEREO-A) dashed vertical lines indicate start and end of an ICME.

As a further example, we show in the bottom panel of Fig. 5 the scenario of an underestimated flow speed. Here, the CH latitude at "L5" is negative ($Lat^{L5} = -35^\circ$), indicating a middle-latitude position, with a large positive B0 angle ($B_0^{L5} = 5.9^\circ$) and a negative B0 angle at "L1" ($B_0^{L1} = -4.8^\circ$). This set of parameters may suggest that the SW observed at STEREO-B (427 km/s in this example) is expected to be increased by the time it reaches STEREO-A (582 km/s in this example). Typically, a larger CH fractional area is expected to produce a larger HSS as suggested in studies correlating CH area with HSS velocity peaks[17]. More recent findings show that this assumed simple linear relationship between the size of CHs and SW speed phenomena is more complex and for very large coronal holes, the relationship starts to saturate[24,26]. However, as illustrated in Fig. 5 (top and bottom panels), a CH with a small fractional area ($A < 0.1$) can generate an HSS with either a higher or lower peak velocity. In our model, this variation arises from the combined effects of the coronal hole (CH) area, latitude, and B0 angle. This suggests that even a small CH can generate a strong high-speed stream (HSS). A useful analogy is a garden hose: an object positioned near the centreline of the water flow receives more water, whereas an object farther away only catches splashes. Similarly, in the context of high-speed streams, a satellite positioned closer to the centreline of the stream measures a higher solar wind velocity, while one farther away records a lower velocity. This effect is intensified for smaller CHs at higher latitudes, leading to greater discrepancies in velocity measurements for large distances between B0 angles. In contrast, larger CHs distribute the solar wind more uniformly, resulting in less variation across different B0 angles.

This pattern is consistent with another study[41], which demonstrated that the HSSs emerging from CHs exhibit a highly structured velocity profile, especially for smaller CHs. The effect becomes more pronounced at higher latitudes, contributing to irregular SW impacts across different regions. A follow-up study further explored this, showing that for larger CHs, there is a saturation effect in the HSS velocity, leading to a more uniform SW distribution as the stream spreads across a wider area[26]. These findings align well with MHD simulations, highlighting the correlation between the size of CHs and the consistency of SW distribution[27].

The identified patterns exhibit consistency across all analysed years and remain robust across different spacecraft configurations. The effect of overestimation or underestimation is more pronounced for CHs with smaller areas, higher latitudes, and significant differences in B0. Therefore, CH latitude and area, along with B0, are key indicators of potential variations in SW velocity between different spacecraft. In the next section, we develop criteria to predict whether HSSs will lead to an increase or decrease in SW velocity at "L1" compared to "L5" and introduce an indicator to estimate the magnitude of these variations.

## Predictive indicator of solar wind velocity variations at "L1" relative to "L5"

As discussed, the combination of CH area, latitude, and B0 angle provides predictive insights into discrepancies in SW velocities at different viewpoints. We now introduce the following indicator to capture these variations:

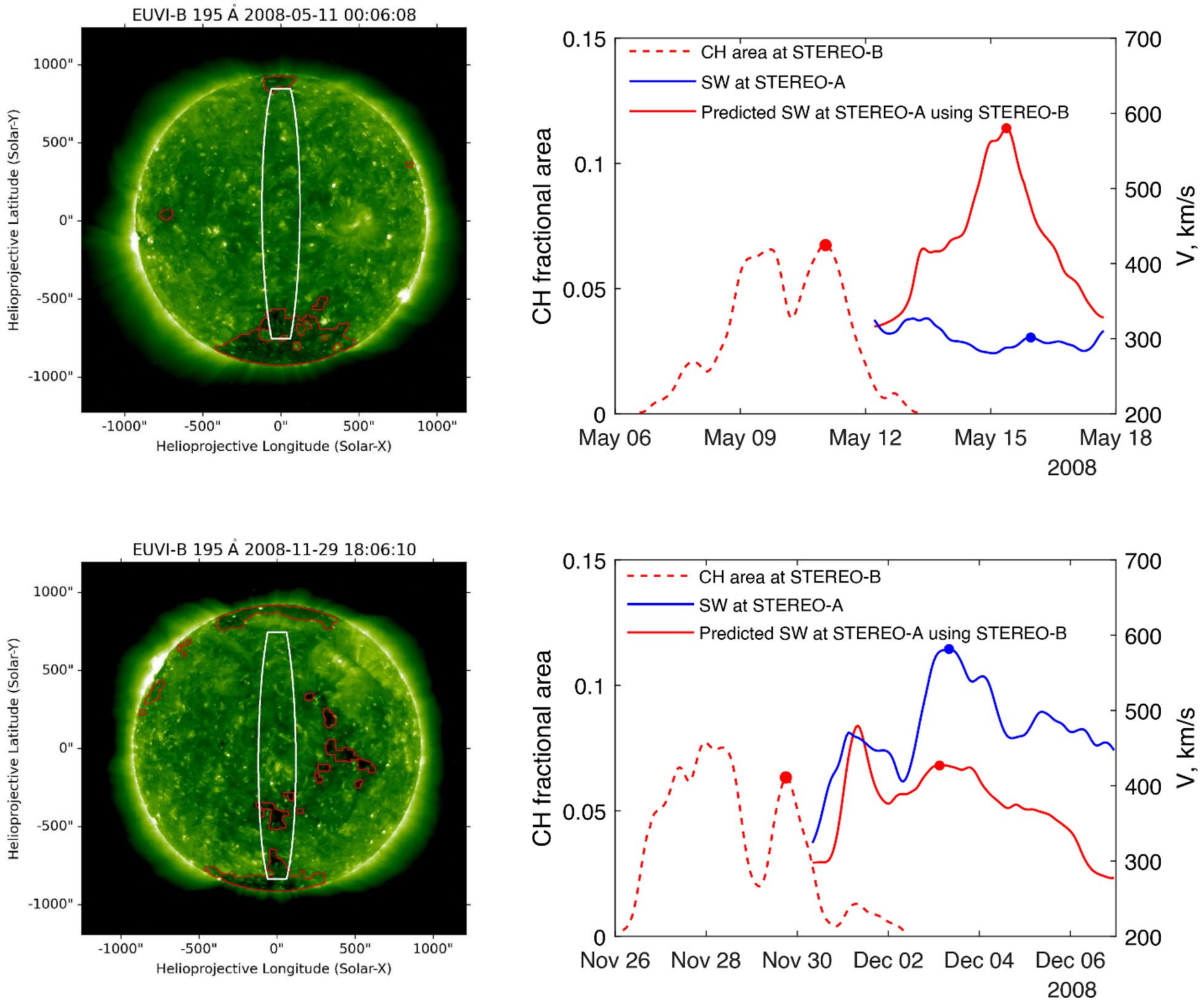

**Fig. 5**. CH and B0 angle impact on L1 – L5 SW velocity. Left panels show 195 Å STEREO-B images together with segmented CHs and $15^{\circ}$-wide meridional slice for 11 May 2008, 00:06 UT (top), and for 29 November 2008, 18:06 UT (bottom). Right panels give overestimated (top) and underestimated (bottom) SW velocity forecasts. The left Y-axis displays the CH area on "L5" (STEREO-B, red dotted line), while the right Y-axis shows the corresponding SW velocity for "L5" (STEREO-B, red) and "L1" (STEREO-A, blue). Blue and red markers indicate the peaks of associated CH-HSS pairs.

$$I = Lat^{L5}(B_0^{L1} - B_0^{L5}) \tag{3}$$

Here, $Lat^{L5}$ denotes the latitude of the CH's centre of mass at "L5", while $B_0^{L5}$ and $B_0^{L1}$ represent the B0 angles at "L5" and "L1", respectively. Let us now consider the following differences in SW velocities:

$$D^V = V^{L1} - V_p^{L5} \tag{4}$$

where, $V_p^{L5}$ is the HSS velocity peak at "L5" and $V^{L1}$ is the mean velocity over the SW velocity segment at "L1" spanning the time between the "L5" and "L1" peaks. The difference $D^V$ at each time step shows the value of discrepancy between "L5" and "L1".

Figure 6a shows the values of $D^V$ ( blue, left Y-axis) together with predictive indicator $I$ (red, right Y-axis). As can be seen from the figure, there is a high dependence ($r = 0.85$) between the blue and red lines. This outcome allows us to use the suggested indicator $I$ to estimate the level of discrepancy between L5 and L1 in advance. For example, CH-HSS No. 13, which exhibits a significant overestimation with a $D^V$ of $-292$ km/s at "L5" compared to "L1" is associated with a large negative $I$ value of $-331$. Conversely, CH-HSS No. 30, characterized by a substantial underestimation with a $D^V$ of 151 km/s, shows a large positive $I$ value of 370. To ensure reliable adjustments to initial predictions, we propose a critical threshold for $I$ to guide decisions on whether the SW velocity at "L1" will increase or decrease relative to "L5" taking into account all spacecraft combinations and years. If any of the following criteria are satisfied, the adjustment will be applied to decrease the SW velocity at "L1" compared to "L5":

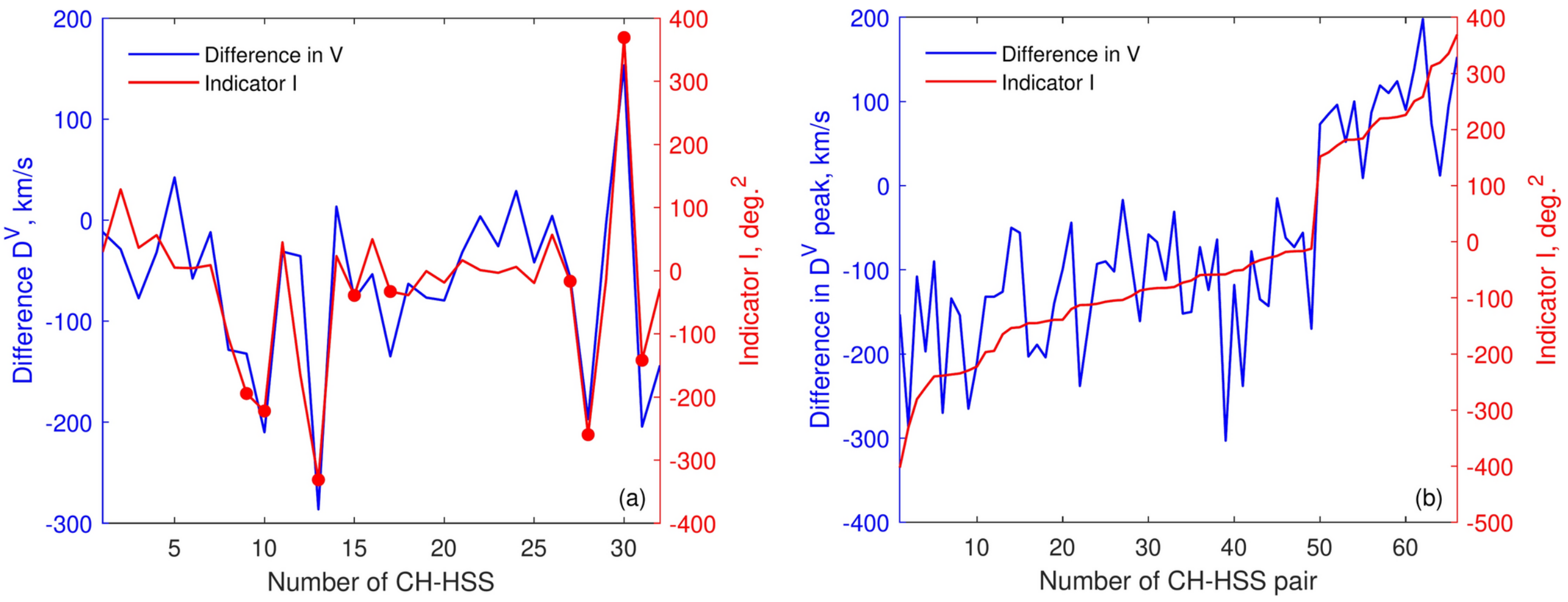

**Fig. 6**. Predictive indicator of SW velocity increase or decrease at "L1" relative to "L5". Panel (a) presents the difference in SW velocity peaks (left Y-axis) and the associated predictive indicator *I* (right Y-axis) for the "L5" (STEREO-B) – "L1" (STEREO-A) 2008 case. Red markers highlight CH-HSS pairs that meet the threshold for the predictive indicator *I*. Panel (b) illustrates the same as panel (a) but only for CH-HSS cases that satisfy the threshold values of the predictive indicator *I* for all considered spacecraft combinations and years.

$$\begin{aligned} (abs(B_0^{L1} - B_0^{L5}) \geq 1)\ \&\ (A_p^{L5} < 0.1)\ \&\ (I \leq -12) \\ (abs(B_0^{L1} - B_0^{L5}) \geq 10)\ \&\ (I \leq -28) \\ (abs(B_0^{L1} - B_0^{L5}) \geq 1)\ \&\ (I \leq -200) \end{aligned} \tag{5}$$

The first criterion in Eq. (5) suggests that discrepancies in SW velocity between "L5" and "L1" are most pronounced for small CHs ($A_p^{L5}$ - the peak of CH area at "L5"). Specifically, the requirement $I \leq -12$ applies to both CHs near the equator and those at higher latitudes. Additionally, a minimal angular distance between B0 angles at "L5" and "L1" is necessary. Conversely, if the angular distance between B0 angles is exceptionally large (second criterion), or if the CH is located at a higher latitude (third criterion), regardless of its size, the HSS is expected to be smaller at "L1" compared to "L5". Thresholds for increase of SW at "L1" compared to "L5" are similar to the last expression in Eq. (5), but with opposite sign for *I*

$$(abs(B_0^{L1} - B_0^{L5}) \geq 1)\ \&\ (I \geq 142) \tag{6}$$

Figure 6b displays the values of the predictive indicator *I* that meet the specified thresholds, along with the corresponding differences $D^V$ for all spacecraft combinations and years considered. The figure clearly shows a strong correlation ($r = 0.83$) between the predictive indicator *I* and the differences $D^V$. To estimate the level of discrepancy in SW velocity between "L5" and "L1" in advance, we use these observed relationships to construct the following third-order linear regression:

$$D^V = \beta_0 + \beta_1 I + \beta_2 I^2 + \beta_3 I^3. \tag{7}$$

Here, the vector of regression coefficients $|\beta_0,\ \beta_1,\ \beta_2,\ \beta_3|^T$ is determined from the least-squares method.

The thick green line in Fig. 7 represents the adjusted SW velocity predictions, which account for the estimated discrepancy between "L5" and "L1" for HSSs where the predictive indicator *I* meets the chosen critical threshold. Initially, we adjust the SW peak at "L5" based on the estimated $D^V$ value (cyan markers). Subsequently, we proportionally adjust the entire HSS profile, from start to end, based on the "L5" observations. As shown in Fig. 7, the correction decisions were made accurately in all cases. For example, the SW prediction for No. 30 was appropriately increased, while for the other cases, it was decreased. Before the correction, the RMS error between SW peaks $V_p^B$ at "L5" and the corresponding values $V^{L1}$ at "L1" was 172 km/s for cases where the predictive indicator signaled a change. After the adjustments, this error significantly decreased to 64 km/s. Overall, the RMS error of the adjusted SW forecast for 3.04 to 6.67 days ahead decreased from 75 to 67 km/s, and the correlation coefficient improved from $r = 0.78$ to $r = 0.83$. This demonstrates that the proposed predictive indicator, which combines CH area, latitude, and B0 angle information, can further enhance the accuracy of SW velocity predictions using L5-L1 observations.

The figures showing the initial and adjusted forecasts for all spacecraft combinations and years are provided in Supplementary Information (Figs. 1-6), with the corresponding forecast statistics summarised in Table 1. The "Set" column lists the spacecraft pairs used for SW predictions, with the spacecraft representing "L5" and "L1" indicated in brackets, signifying that predictions are made from "L5" to "L1". The "Year" column specifies the year of the data used, while the "Lead Time" column shows the prediction lead time in days. The "Initial Prediction" column presents the RMS error (in km/s) and correlation coefficient ($r$) between the true and predicted data, considering the separation angle between the spacecraft and their distances from the Sun. Additionally, we include the RMS error ($RMS_p$) specifically for the HSS peaks where the predictive indicator signaled the discrepancies between SW velocity at "L5" compared to "L1". The "Adjusted Prediction" column provides the statistics after correcting for these discrepancies.

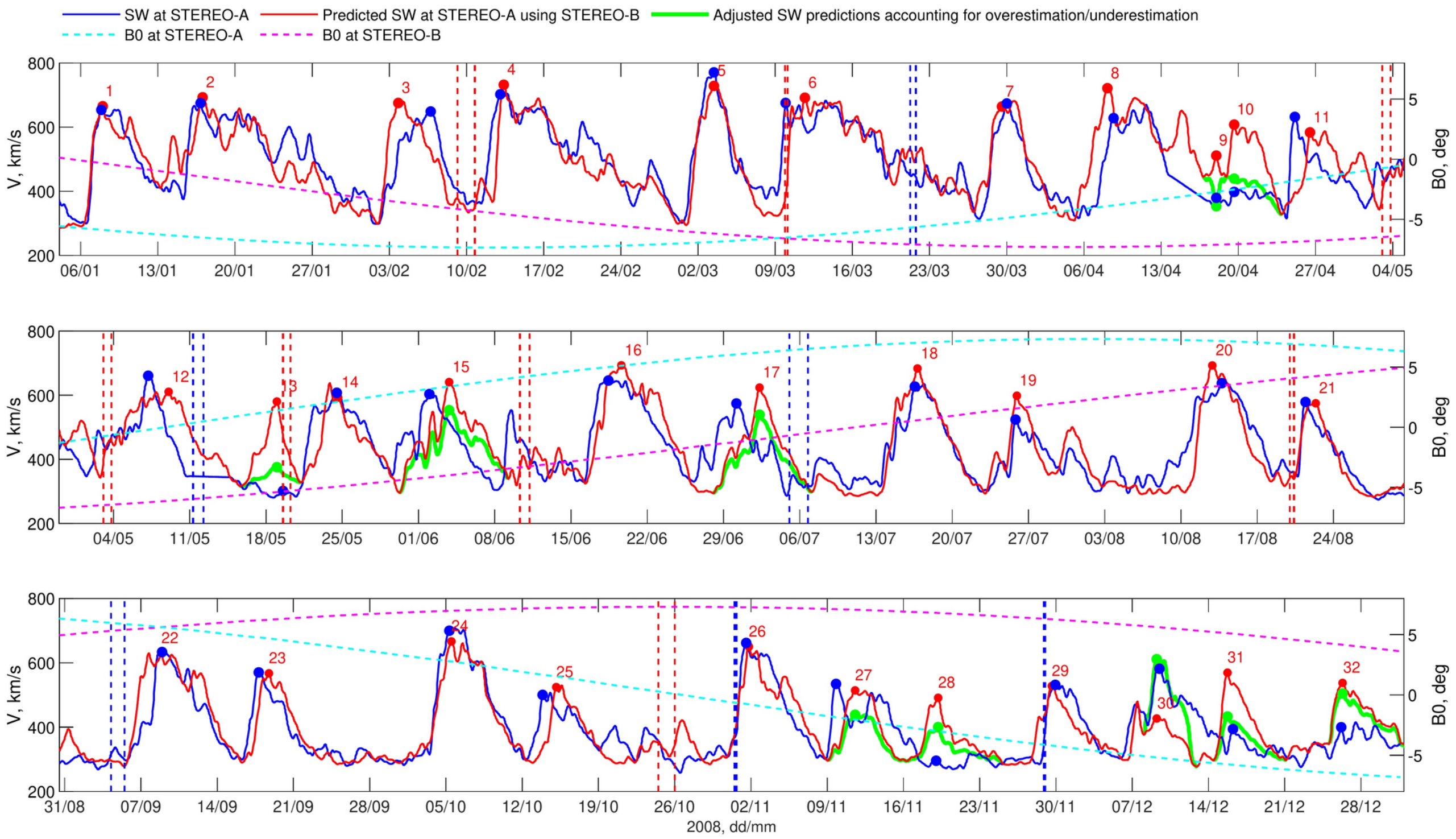

**Fig. 7**. The same as Fig. 4, but with the adjusted prediction of SW velocities, accounting for overestimation or underestimation (thick green lines).

| Set | Year | Lead time in days | Initial prediction | | | Adjusted prediction | | |
|---|---|---|---|---|---|---|---|---|
| | | | *RMS* | $RMS_P$ | *r* | *RMS* | $RMS_P$ | *r* |
| B(L5)–A(L1) | 2008 | 3.04–6.67 | 75 | 172 | 0.78 | 67 | 64 | 0.83 |
| B(L5)–E(L1) | 2008 | 1.54–3.45 | 58 | 115 | 0.88 | 55 | 39 | 0.89 |
| E(L5)–A(L1) | 2008 | 1.51–3.23 | 63 | 162 | 0.86 | 57 | 76 | 0.87 |
| B(L5)–E(L1) | 2009 | 3.23–5.14 | 60 | 101 | 0.55 | 56 | 39 | 0.60 |
| E(L5)–A(L1) | 2009 | 3.14–4.86 | 60 | 129 | 0.57 | 53 | 85 | 0.67 |
| B(L5)–E(L1) | 2010 | 4.84–6.79 | 90 | 173 | 0.52 | 79 | 60 | 0.63 |
| E(L5)–A(L1) | 2010 | 4.76–6.50 | 86 | 139 | 0.53 | 83 | 74 | 0.56 |

**Table 1**. Forecast statistics for all the considered combinations of spacecraft and years. "A" - STEREO-A, "B" - STEREO-B, "E" - Earth.

As shown in Table 1, the RMS error of SW predictions at "L1" using data from "L5" after accounting for discrepancies between "L5" and "L1", ranges from 55 to 83 km/s for lead times between 1.51 and 6.79 days. The introduced predictive indicator enables estimation of these discrepancies based solely on CH latitude and B0 angle information. We note, that the most accurate results are achieved when predicting SW velocity at STEREO-A using data from STEREO-B for 2008, with lead times between 3.04 and 6.67 days. The larger lead time in the STEREO-B to STEREO-A 2008 case increases spatial separation, which naturally introduces more uncertainty as solar wind structures evolve. However, it achieves an RMS error of 75 km/s and $r = 0.78$ over 3.04 – 6.67 days, outperforming the STEREO-B to Earth prediction in 2008 (RMS error of - 58 km/s, $r = 0.88$), which benefits from a shorter separation (1.54 – 3.45 days) and less evolved solar wind structures. When considering RMSE relative to lead time and separation distance, the Stereo-B to Stereo-A constellation shows a stronger performance. The variation in statistical accuracy may be influenced by factors such as uncertainties in CH data, biased time shifts, overlaps with CMEs, and differences in instrument calibration - areas that warrant further investigation for both scientific and operational purposes.

## Conclusions
This study presents meaningful advancements in our understanding of the relationship between HSSs and their solar sources - CHs - by linking their position and characteristics to SW dynamics, through the simulation of an L5-L1 observational configuration. We identified that the CH latitude, when combined with the B0 angle, is a critical factor influencing discrepancies in SW velocities between the L5 and L1 Lagrange points, particularly for small CHs. This effect is intensified for smaller CHs located at higher latitudes, leading to greater discrepancies for large B0 angle distances, whereas larger CHs produce more uniform SW impacts across both locations.

Our analysis also demonstrates the high potential of L5 observations for enhancing SW predictions at L1, achieving reasonable accuracy with RMS errors between 55 and 83 km/s and lead times extending up to nearly seven days. The introduction of a predictive indicator and empirical criteria based on CH latitude, area, and B0 angle, which account for discrepancies in SW velocities at L5 and L1, refines these predictions. This sets an extraordinary benchmark for future mission designs with enhanced monitoring capacity, ultimately providing more reliable space weather tools.

Moreover, this research underscores the importance of integrating CH location and area information from instruments like the EUVI imager onboard the Vigil mission into L5 missions. Such integration will not only improve space weather forecasting but also advance our fundamental understanding of the solar-terrestrial environment, contributing to the broader field of heliophysics. These outcomes highlight the value of continued exploration and observation from diverse vantage points in space, such as L5, L4, and others, to fully unravel the complexities of our Sun's influence on the Solar System.

## Data availability
The data were obtained from the STEREO/PLASTIC database at https://stereo-ssc.nascom.nasa.gov/data/ins_data/plastic/level2/Protons/Derived_from_On_Board_Calculation/ and the OMNI-2 database at https://omniweb.gsfc.nasa.gov/form/dx1.html, along with interplanetary coronal mass ejections (ICMEs) observed with STEREO-A and STEREO-B at https://stereo-ssc.nascom.nasa.gov/pub/ins_data/impact/level3/LanJian_STEREO_ICME_List.txt, as well as the Richardson & Cane ICME list at https://izw1.caltech.edu/ACE/ASC/DATA/level3/icmetable2.htm for Earth. CH data will be shared on reasonable request to the corresponding author.

## Acknowledgements

The authors acknowledge the financial support by the University of Graz and the ESA SSA P3-SWE-IV.2: "Use of L5 data in CME propagation models". The authors acknowledge The STEREO/SECCHI data are produced by an international consortium of the Naval Research Laboratory (USA), Lockheed Martin Solar and Astrophysics Lab (USA), NASA Goddard Space Flight Center (USA), Rutherford Appleton Laboratory (UK), University of Birmingham (UK), Max-Planck-Institut für Sonnenforschung (Germany), Centre Spatiale de Liége (Belgium), Institut d'Optique Théorique et Appliquée (France), and Institut d'Astrophysique Spatiale (France). We thank the referees for valuable comments on this study.

## Author contributions

T.P. developed the method and led the writing of the paper. A.V. and M.T. contributed to the conceptualisation of the study, analysis, and writing. S.J.H. performed CH detection, extracted CH parameters, and contributed to the analysis. All authors discussed the results and provided feedback on the manuscript.

## Declarations

## Competing interests

The authors declare no competing interests.

## Additional information

**Supplementary Information** The online version contains supplementary material available at https://doi.org/10.1038/s41598-025-97246-2.

**Correspondence** and requests for materials should be addressed to T.P.